\pdfoutput=0
\documentclass[journal]{IEEEtran}[10pt]
\usepackage{amsfonts}
\usepackage{graphicx}
\usepackage{color}
\usepackage{amsmath,amsfonts,amssymb,amsthm,epsfig,epstopdf,url,array}
\usepackage{url,textcomp}
\usepackage{authblk}
\usepackage{cite}
\usepackage{caption}
\usepackage{float}
\usepackage[misc]{ifsym}
\usepackage[breaklinks,colorlinks,linkcolor=black,citecolor=black,urlcolor=black]{hyperref}

\newtheorem{theorem}{Theorem}

\newcommand\blfootnote[1]{%
  \begingroup
  \renewcommand\thefootnote{}\footnote{#1}%
  \addtocounter{footnote}{-1}%
  \endgroup
}
\begin{document}
\title{Outage Performance Analysis of Type-I HARQ Aided V2V Communications}

\author{\IEEEauthorblockN{Huan Zhang\IEEEauthorrefmark{1}, Zhengtao Liao\IEEEauthorrefmark{4}, Zheng~Shi$^{\textrm{\Letter}}$\IEEEauthorrefmark{2}\IEEEauthorrefmark{3},Guanghua~Yang\IEEEauthorrefmark{2},  Qingping Dou\IEEEauthorrefmark{2}, and  Shaodan~Ma\IEEEauthorrefmark{3}\\
%\vspace{-1em}
\IEEEauthorrefmark{1}Xiaomi Communication Technology Co., Ltd., Beijing 100085, China\\
\IEEEauthorrefmark{2}School of Intelligent Systems Science and Engineering, Jinan University, Zhuhai, China\\
\IEEEauthorrefmark{3}Department of Electrical and Computer Engineering, University of Macau, China}\\
\IEEEauthorrefmark{4} School of Electronic and Information Engineering, South China University of Technology, Guangzhou, China
\vspace{-1.5em}}

%\affil{}
\maketitle
\begin{abstract}
   Vehicle-to-vehicle (V2V) communications under dense urban environments usually experience severe keyhole fading effect, especially for multi-input multi-output (MIMO) channels, which degrades the capacity and outage performance due to the rank deficiency. To avoid these, the integration of MIMO and Type-I hybrid automatic repeat request (HARQ) is proposed to assist V2V communications in this paper. By using the Meijer G-function, the outage probability of the proposed V2V communications system is derived in closed form. With the result, meaningful insights are gained by conducting the asymptotic outage analysis. Specifically, it is revealed that full-time diversity order can be achieved, while full spatial diversity order is unreachable as compared to Type-I HARQ aided MIMO systems without keyhole effect. Moreover, we prove that the asymptotic outage probability is a monotonically increasing and convex function of the transmission rate. Finally, the analytical results are validated through extensive numerical experiments.

\end{abstract}
% Note that keywords are not normally used for peer review papers.
\IEEEpeerreviewmaketitle
\hyphenation{HARQ}
\section{Introduction}\label{sec:int}
\blfootnote{This work was supported in part by Guangdong Basic and Applied Basic Research Foundation under Grant 2019A1515012136, in part by National Natural Science Foundation of China under Grants 62171200, and
62171201, in part by Zhuhai Basic and Applied Basic Research Foundation under Grant ZH22017003210050PWC, in part by 2018 Guangzhou Leading Innovation Team Program (China) under Grant 201909010006, in part by
the Science and Technology Development Fund, Macau SAR (File no. 0036/2019/A1 and File no. SKL-IOTSC(UM)-2021-2023), and in part by the Research Committee of University of Macau under Grant MYRG2020-00095-
FST.}
\blfootnote{\emph{Corresponding author: Zheng Shi (E-mail: zhengshi@jnu.edu.cn).}}

%\subsection{Background}
\IEEEPARstart{W}{ith} the rapid development of intelligent transportation systems, vehicle-to-vehicle (V2V) communications have been widely studied in recent years. Specifically, V2V communications only allow the exchange of information between adjacent vehicles with short distance,  which enhances the transmission reliability, supports delay-sensitive applications, and improves traffic safety \cite{jie2018latency}.  However, V2V communications considerably differ from mobile cellular communications. On the one hand, both transmit and receive vehicles are in motion, which results in more significant Doppler effects and more rapid channel dynamics than cellular communications. On the other hand, the transceiver antennas are mounted at almost the same height, where the local scatterers in the surrounding environment incur more than one multiplicative small-scale fading processes, i.e., cascaded fading \cite{alghorani2020improved}. The local scattering objects include buildings, vehicles, street corners, tunnels, etc., which obstruct the direct link between two vehicles and lead to non-line-of-sight (NLOS) propagation channel condition \cite{matolak2011worse,liang2019large} (cf. Fig. 1). To characterize cascaded fading in V2V communications, the authors in \cite{ali2018throughput,fu2018applying,bithas2017double} proposed to use double-bounce/multiple scattering distributions, such as double-Rayleigh, double-Nakagami-$m$, double-Weibull, and double-generalized Gamma distributions. %Moreover, a decode-and-forward relaying scheme was developed in \cite{bithas2017v2v} for V2V communications by considering double-Nakagami fading, where both the exact and asymptotic outage probabilities were derived. It was demonstrated experimentally in \cite{matolak2011worse} that the double-bounce scattering distributions can provide an accurate statistical fit for channel modeling of V2V communications.
%However, both experimental and theoretical results verified that the error performance of the multiple scattering model is worse than that of the traditional Rayleigh channel model for cellular communication systems. Therefore, V2V communications often suffer from a more severe fading leading to lower spectral efficiency and reception reliability than cellular communications.

To boost the spectral efficiency and reliability, multi-input multi-output (MIMO) aided V2V communications have drawn an ever-increasing attention, because multiple antennas can be easily placed on vehicles with large surface \cite{yang2021asymptotic,xu2021sum,zhang2021large,gong2021unified}. {\color{black}In contrast to single-input single-output (SISO) systems, MIMO systems equipping with multiple antennas are capable of reaping the benefit of spatial multiplexing gain.} Nevertheless, in realistic propagation environments, the performance of MIMO systems is also susceptible to multiple scattering propagation. {\color{black}Particularly for MIMO assisted V2V (MIMO-V2V) communications, multiplicative fading processes encountered in multiple scattering condition are inevitable due to the mobility of the vehicles and low elevation height of the transceiver antennas\cite{wu2020tensor}. In dense urban environments, all the MIMO-V2V propagation paths travel through the same narrow pipe, which results in the so-called keyhole effect \cite{alghorani2017on}.}
However,  most of the relevant works, i.e., \cite{matolak2011worse,chizhik2002keyholes,ngo2017no,shin2003capacity}, have demonstrated that keyhole channels not only offset the advantage of spatial diversity of MIMO, but also degrade the spatial multiplexing gain. Since the keyhole effect negatively impacts the reliability of MIMO-V2V communications, it is of necessity to remedy the performance loss for MIMO-V2V communications.

To address the above issue, hybrid automatic repeat request (HARQ) is a promising technique to support reliable communications \cite{shi2019effective,wang2020outage}. Specifically, the essence of HARQ is utilizing both forward error control and automatic repeat request \cite{dahlman20103g}. Based on whether the erroneously received packets are discarded or not, and what types of coding and decoding strategies are used, HARQ can be further divided into the following three basic schemes, namely, Type-I HARQ, HARQ with chase combining (HARQ-CC) and HARQ with incremental redundancy (HARQ-IR). In contrast to the HARQ-CC and HARQ-IR schemes, the Type-I HARQ scheme performs the decoding based on the currently received packet without storing the failed
packets, which has been widely adopted to assist wireless communications \cite{shi2015outage}. Nevertheless, the performance of HARQ schemes over keyhole fading channels was seldom reported in the literature.

In order to extend the application of MIMO to V2V communications, this paper focuses on the performance investigation of Type-I HARQ aided V2V communications over MIMO keyhole fading channels. Firstly, the exactly closed-form expression is derived for the outage probability of  Type-I HARQ  assisted V2V communications with MIMO keyhole effect. Moreover, the asymptotic outage probability is also conducted. With asymptotic result, it is concluded that full spatial diversity order is unreachable with MIMO, while full time diversity order can be achieved from using Type-I HARQ. This obviously justifies the effectiveness of using Type-I HARQ to conquer keyhole effect. More specifically, the spatial diversity order is determined by the minimum of the numbers of transmit and receive antennas. Moreover, it is proved that the asymptotic outage probability is a monotonically increasing and convex function with respect to (w.r.t.) transmission rate. This property facilitates the optimal rate selection for practical system design.

% \emph{Notation}: The following notations will be used throughout this paper. Bold uppercase and lowercase letters are used to denote matrices and vectors, respectively. ${\bf X}^{\mathrm{H}}$, ${\bf X}^{-1}$, ${\mathrm{det}}({\bf X})$ and ${\rm tr}({\bf X})$ stand for the conjugate transpose, the inverse, the  determinant and the trace of matrix ${\bf X}$, respectively. $\mathbf{I}$ represents an identity matrix. $\left\| \cdot \right\|$ denotes the Euclidean norm of a vector. ${\mathcal {CN}}({\bf{0}},{\bf I})$ represents the  complex Gaussian vector with zero mean vector and identity covariance matrix. $\mathbf{A}\succ\mathbf{B}$ means that $\mathbf{A}-\mathbf{B}$ is a positive definite matrix. ${\rm i}=\sqrt{-1}$ denotes the imaginary unit. The symbol ``$\simeq$'' denotes ``asymptotically equal to''.
% The definitions of any other notations are deferred to the place where they arise.
\section{System Model}\label{sec:sys_mod}
\begin{figure}
  \centering
  \includegraphics[width=3.5in]{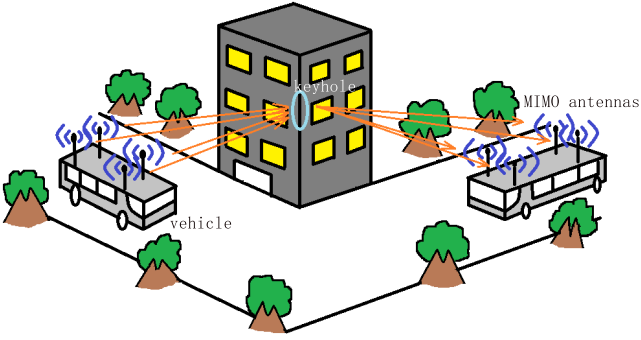}
  \caption{An example for HARQ-MIMO assisted V2V communications with keyhole effect.}\label{fig:model}
\end{figure}
As shown in Fig. \ref{fig:model}, we consider a V2V communication system under urban environments where the transmit vehicle and the receive vehicle are equipped with $N_{T}$ and $N_{R}$ antennas, respectively. We assume that there are a number of obstacles between the transmitter and receiver, and the transmitted signal propagates through electromagnetically small apertures (or keyholes) among obstacles. By following the MIMO keyhole channel model in \cite{shin2003effect}, the channel matrix ${\bf H}$ for V2V communications is given by%$N_R\times N_T$
\begin{equation}\label{eqn:channel_model}
{{\bf{H}}} = {{\bf{u}}}{{\bf{v}}}^{\rm{H}} = \left( {\begin{array}{*{20}{c}}
{{u_{1}}{v_{1}^\ast}}&{{u_{1}}{v_{2}^\ast}}& \cdots &{{u_{1}}{v_{N_T}^\ast}}\\
{{u_{2}}{v_{1}^\ast}}&{{u_{2}}{v_{2}^\ast}}& \cdots &{{u_{2}}{v_{N_T}^\ast}}\\
 \vdots & \vdots & \ddots & \vdots \\
{{u_{{N_R}}}{v_{1}^\ast}}&{{u_{{N_R}}}{v_{2}^\ast}}& \cdots &{{u_{{N_R}}}{v_{N_T}^\ast}}
\end{array}} \right),
\end{equation}
where each entry of ${\bf{v}}$ and ${\bf{u}}$ follows independent and identically distributed (i.i.d.) complex normal distribution, i.e., ${\bf{v}} \sim {\mathcal {CN}}({\bf{0}},{\bf I}_{N_T})$ and ${\bf{u}} \sim {\mathcal {CN}}({\bf{0}},{\bf I}_{N_R})$.

Clearly from \eqref{eqn:channel_model}, ${{\bf{H}}}$ is a rank-one matrix in the presence of the keyhole effect. The rank-one deficiency significantly reduces the spatial multiplexing gain and degrades the MIMO capacity. To enhance the reception reliability, the  Type-I HARQ is employed in this paper. Therefore, the accumulated mutual information obtained by Type-I HARQ scheme after $K$ HARQ rounds can be expressed as \cite{7438859}
\begin{equation}\label{eqn:mu}
 I = \max\limits_{k=1,\cdots,K}{\log _2}\left( {\det \left[ {\mathbf{I}_{{N_R}}} + \frac{{{\gamma _k}}}{{{N_T}}}{{\bf{H}}_k}{{\bf{H}}_k}^{\rm{H}} \right]} \right),
\end{equation}
where ${\bf{H}}_k,~k\in\{1,\cdots,K\}$ represents the keyhole channel matrix for the $k$-th HARQ transmission round and is modelled according to (\ref{eqn:mu}), and  $\gamma _k$ stands for the average transmit SNR at the transmitter for the $k$-th round. Besides, ${\bf{H}}_1, \cdots, {\bf{H}}_K$ are assumed to be independent random matrices.

Clearly from (\ref{eqn:channel_model}), since each entry of the channel matrix is expressed as a product of two complex Gaussian random variables, this complex form hinders the following outage analyses. Besides, the accumulated mutual information for the Type-I HARQ scheme involves many matrix operations, such as determinant operations, product of block matrix, which further complicates the analysis. To the best of our knowledge, there is no available results for the outage performance of Type-I HARQ  aided V2V communications with MIMO keyhole effect in the literature.
\section{Analysis of Exactly Outage Probability}\label{sec:out}
To investigate the performance of Type-I HARQ aided V2V communication systems, the outage probability is the most significant and essential performance metric. More specifically, the outage probability is defined as the probability of the event that the accumulated mutual information is less than the preset transmission rate $R$. According to the definition of outage probability and combining with MIMO keyhole channel model (\ref{eqn:channel_model}), the outage probability of the proposed systems can be rewritten as
 \begin{align}\label{eqn:out_probI}
&P_{out}\notag\\
&=\Pr \left( {\max\limits_{k=1,...,K}{\log _2}\left( {\det \left[ {\mathbf{I}_{{N_R}}} + \frac{{{\gamma _k}}}{{{N_T}}}{{\bf{H}}_k}{{\bf{H}}_k}^{\rm{H}} \right]} \right) < R} \right) \notag\\
&= \prod\limits_{k = 1}^K {\Pr \left( {{{\log }_2}\left( {1 + \frac{{{\gamma _k}}}{{{N_T}}}{{\left\| {{{\bf{u}}_k}} \right\|}^2}{{\left\| {{{\bf{v}}_k}} \right\|}^2}} \right) < R} \right)} \notag\\
&=\prod\limits_{k = 1}^K {{F_{{X_k}}}\left( {\frac{{{N_T}}}{{{\gamma _k}}}\left( {{2^R} - 1} \right)} \right)},
\end{align}
where the second step holds by using $\mathrm{det}(\mathbf{I}+\mathbf{A}\mathbf{B})=\mathrm{det}(\mathbf{I}+\mathbf{B}\mathbf{A})$, and  ${F_{{X_k}}}(x)$ denotes the cumulative distribution function
(CDF) of $X_k = {{\left\| {{{\bf{u}}_k}} \right\|}^2}{{\left\| {{{\bf{v}}_k}} \right\|}^2}$. From (\ref{eqn:out_probI}), the outage analysis of Type-I HARQ scheme boils down to determining the distribution of the equivalent channel gain ${{\left\| {{{\bf{u}}_k}} \right\|}^2}{{\left\| {{{\bf{v}}_k}} \right\|}^2}$. Since ${\left\| {{{\bf{u}}_k}} \right\|}^2$ and ${\left\| {{{\bf{v}}_k}} \right\|}^2$ are central chi-square random variables with $2N_R$ and $2N_T$ degrees of freedom, respectively. The PDF of $X_k$ is given as \cite{shin2003capacity}
\begin{align}\label{eqn:pdf_X_l}
{f_{{X_k}}}\left( x \right)=  \frac{{2{x^{\left( {{N_T} + {N_R}} \right)/2 - 1}}{K_\tau}\left( {2\sqrt x } \right)}}{{\Gamma \left( {{N_T}} \right)\Gamma \left( {{N_R}} \right)}},
\end{align}
 where $\Gamma(\cdot)$ is the gamma function \cite[Eq. (8.310)]{jeffrey2007table}, $K_{\tau}(\cdot)$ represents the modified
Bessel function of the $\tau$-th order \cite[Eq. (8.432.6)]{jeffrey2007table}, and $\tau=\left| {{N_T} - {N_R}} \right|$. In analogy to \cite{adamchik1990algorithm}, it is suggested to invoke Meijer G-function to generalize our analytical results. By utilizing \cite[Eq. (9.34.3)]{jeffrey2007table}, the PDF of $X_k$ can be expressed in the form of Meijer G-function as \cite[Eq. (9.301)]{jeffrey2007table}
\begin{align}\label{eqn:pdf_X_l_Meij}
{f_{{X_k}}}\left( x \right) &= \frac{{G_{0,2}^{2,0}\left( {\left. {\begin{array}{*{20}{c}}
 - \\
{{N_T},{N_R}}
\end{array}} \right|x} \right)}}{{x\Gamma \left( {{N_T}} \right)\Gamma \left( {{N_R}} \right)}}.
\end{align}
Then, the CDF of $X_k$ can be derived by using \cite[Eq. (26)]{adamchik1990algorithm} as
\begin{equation}\label{eqn:CDF_Zl}
{F_{{X_k}}}\left( x \right) =\int_0^x {{f_{{X_k}}}\left( y \right)dy}= \frac{{G_{1,3}^{2,1}\left( {\left. {\begin{array}{*{20}{c}}
1\\
{{N_T},{N_R},0}
\end{array}} \right|x} \right)}}{{\Gamma \left( {{N_T}} \right)\Gamma \left( {{N_R}} \right)}}.
\end{equation}
Finally, substituting  (\ref{eqn:CDF_Zl}) into (\ref{eqn:out_probI}) yields the closed-form expression of outage probability for the MIMO-Type-I HARQ scheme as
\begin{align}\label{eqn:op_type_I}
P_{out}
=&\frac{1}{\left({\Gamma \left( {{N_T}} \right)\Gamma \left( {{N_R}} \right)}\right)^K}\notag\\
\times&\prod\limits_{k = 1}^K {G_{1,3}^{2,1}\left( {\left. {\begin{array}{*{20}{c}}
1\\
{{N_T},{N_R},0}
\end{array}} \right|{\frac{{{N_T}}}{{{\gamma _k}}}\left( {{2^R} - 1} \right)}} \right)}.
\end{align}

\section{Analysis of Asymptotic Outage Probability}\label{sec:out_asy}
However, although the Meijer G-function in (\ref{eqn:op_type_I}) is a built-in function in many popular mathematical software packages, such as MATLAB, the integral form of Meijer G-function are still complex which hampers the extraction of useful insights, such as diversity order and coding gain. To overcome this issue, an asymptotic expression of the asymptotic outage probability needs to be obtained in the high SNR regime

It is obviously that the asymptotic behavior of (\ref{eqn:op_type_I}) under high SNR, i.e., $\gamma_k\to\infty$, corresponds to the behavior of the PDF (\ref{eqn:pdf_X_l}) at small $x$, i.e., $x\to 0$. Based on \cite[Eq. (10.30.2), Eq. (10.30.3)]{lozier2003nist}, the modified Bessel function for $x\to 0$ is asymptotic to
 \begin{equation}\label{eqn:asyp_Kv}
{K_\tau}\left( x \right) \simeq \left\{ {\begin{array}{*{20}{c}}
{\frac{1}{2}\Gamma \left( \tau \right){{\left( {\frac{1}{2}x} \right)}^{ - \tau}},}&{\tau > 0},\\
{ - \ln \left( x \right)},&{\tau = 0}.
\end{array}} \right.
\end{equation}
 By substituting the asymptotic expression (\ref{eqn:asyp_Kv}) into (\ref{eqn:pdf_X_l}), the asymptotic outage probability for MIMO-Type-I HARQ scheme can be written as
 \begin{align}
&P_{out}^{asy} \notag\\
= &\left\{ {\begin{array}{*{20}{c}}
\prod\limits_{k = 1}^K {\int\limits_0^{\frac{{{N_T}}}{{{\gamma _k}}}\left( {{2^R} - 1} \right)} {\frac{{2{x^{\left( {{N_T} + {N_R}} \right)/2 - 1}}\frac{1}{2}\Gamma \left( {\tau} \right){{\left( {\frac{1}{2}2\sqrt x } \right)}^{ - \tau}}}}{{\Gamma \left( {{N_T}} \right)\Gamma \left( {{N_R}} \right)}}dx} }, &{\textup{$\tau > 0$}},\\
{\left( { - 1} \right)^K}\prod\limits_{k = 1}^K {\int\limits_0^{\frac{{{N_T}}}{{{\gamma _k}}}\left( {{2^R} - 1} \right)} {\frac{{2{x^{{N_T} - 1}}\ln \left( {2\sqrt x } \right)}}{{{\Gamma ^{\rm{2}}}\left( {{N_T}} \right)}}dx} },&{\textup{$\tau=0$}}.
\end{array}} \right.
\end{align}
Accordingly, the following two cases are treated individually.
\subsection{$\tau > 0$}
For the case of $\tau > 0$, we have
  \begin{align}
    &P_{out}^{asy}=\prod\limits_{k = 1}^K {\frac{{\Gamma \left( {\tau} \right)}{\left( {\frac{{{N_T}}}{{{\gamma _k}}}\left( {{2^R} - 1} \right)} \right)}^{\left( {{N_T} + {N_R}} \right)/2 -\tau/2}}{{\Gamma \left( {{N_T}} \right)\Gamma \left( {{N_R}} \right)}\left({\left( {{N_T} + {N_R}} \right)/2 - \tau/2}\right)}}.
 \end{align}
\subsection{$\tau = 0$} For the case of $\tau=0$, one has
  \begin{align}\label{eqn:27}
    P_{out}^{asy}=&
{\left( { - 1} \right)^K}\prod\limits_{k = 1}^K \frac{{2{{\left( {\frac{{{N_T}}}{{{\gamma _k}}}\left( {{2^R} - 1} \right)} \right)}^{{N_T}}}}}{{{\Gamma ^{\rm{2}}}\left( {{N_T}} \right)}}\notag\\
&\times \int\limits_0^1 {{x^{{N_T} - 1}}\ln \left( {2\sqrt {\frac{{{N_T}}}{{{\gamma _k}}}\left( {{2^R} - 1} \right)x} } \right)dx} \notag\\
\mathop \simeq \limits^{\left( a \right)}& \prod\limits_{k = 1}^K {\frac{{{{\left( {\frac{{{N_T}}}{{{\gamma _k}}}\left( {{2^R} - 1} \right)} \right)}^{{N_T}}}\ln \left( {{\gamma _k}} \right)}}{{{\Gamma ^{\rm{2}}}\left( {{N_T}} \right)}}\int\limits_0^1 {{x^{{N_T} - 1}}dx} } \notag\\
 =& \prod\limits_{k = 1}^K {\frac{{{{\left( {\frac{{{N_T}}}{{{\gamma _k}}}\left( {{2^R} - 1} \right)} \right)}^{{N_T}}}\ln \left( {{\gamma _k}} \right)}}{{{N_T}{\Gamma ^{\rm{2}}}\left( {{N_T}} \right)}}},
\end{align}
where step (a) holds by using $-2\ln \left( {2\sqrt {{{{N_T}}}/{{\gamma _k}}\left( {{2^R} - 1} \right)x} } \right)\simeq \ln\left( {{\gamma _k}} \right)$.

As a consequence, the asymptotic outage probability for the Type-I HARQ scheme over MIMO keyhole channel is derived as shown in the following theorem.
\begin{theorem}\label{The:asy_I}
In the high SNR regime, i.e., $\gamma_k\to\infty$, the asymptotic outage probability of the proposed systems with MIMO keyhole effect is given by \eqref{eqn:asy I} at the top of the next page.
\begin{figure*}[!t]
  \begin{align}\label{eqn:asy I}
    P_{out}^{Type-I}\simeq \left\{ {\begin{array}{*{20}{c}}
    {\prod\limits_{k = 1}^K {\frac{{{{\left( {\frac{{{N_T}}}{{{\gamma _k}}}\left( {{2^R} - 1} \right)} \right)}^{{N_T}}}\ln {{\gamma _k}} }}{{{N_T}({{\Gamma}\left( {{N_T}} \right)}) ^{2}}}},}&{\textup{$N_T - N_R = 0$}},\\
    {\prod\limits_{k = 1}^K {\frac{{\Gamma \left( {\tau} \right)}}{{\Gamma \left( {{N_T}} \right)\Gamma \left( {{N_R}} \right)}}\frac{{{{\left( {\frac{{{N_T}}}{{{\gamma _k}}}\left( {{2^R} - 1} \right)} \right)}^{\left( {{N_T} + {N_R}} \right)/2 -\tau/2}}}}{{\left( {{N_T} + {N_R}} \right)/2 - \tau/2}}},}&{\textup{$N_T - N_R \ne 0$}}.\\
    \end{array}} \right.
    \end{align}
  \hrulefill
  \end{figure*}
\end{theorem}

\section{Discussions of Asymptotic Results}\label{sec:div}
To facilitate our discussion, we assume equal transmit SNRs, i.e., $\gamma_1=\cdots=\gamma_K=\gamma$. By identifying the asymptotic results (\ref{eqn:asy I}) with \cite[eq.(3.158)]{tse2005fundamentals}, \cite[eq.(1)]{wang2003simple}, the asymptotic outage probabilities of the proposed systems  with MIMO keyhole effect as $\gamma\to \infty$ can be generalized as \cite{shi2017asymptotic}
\begin{equation}\label{eqn:pro}
P_{out}^{asy}=\left(\mathcal{C}(R)\gamma\right)^{-d}(\ln \gamma)^{K\mathbb I(N_t-N_r)}+{o}(\gamma^{-d}(\ln \gamma)^{K\mathbb I(N_t-N_r)}),
\end{equation}
where $\mathbb I(N_t-N_r)$ denotes the indicator function, $\mathcal{C}(R)$ represents the modulation and coding gain, $d$ stands for the diversity order, and $o(\gamma^{-d})$ denotes the higher order terms. In what follows, the diversity order $d$, and the modulation and coding gain $\mathcal{C}(R)$ of the Type-I HARQ aided V2V communications is individually discussed.
\subsection{Diversity Order}\label{sec:div_order}
The diversity order measures the degrees of freedom of communication systems, which is defined as the declining slope of the outage probability with regard to the transmit SNR on a log-log scale in the high SNR regime as \cite{chelli2014performance}
\begin{equation}\label{eqn:div}
d =-\lim\limits_{\gamma\to\infty}\frac{{\log}\left(P_{out}\right)}{{\log}\left(\gamma\right)}.
\end{equation}
By substituting the  \eqref{eqn:asy I} into (\ref{eqn:div}), we can obtain the diversity order is given by $d=K\mathrm{min}\left(N_T,N_R\right)$, where $K$ and $\mathrm{min}(N_T,N_R)$ represent the achievable time and spatial diversity order, respectively. As proved in \cite{yang2021asymptotic}, the diversity order of MIMO systems without keyhole effect is $N_TN_R$. However, the spatial diversity order of the Type-I HARQ with MIMO keyhole effect is the minimum of the numbers of transmit and receive antennas. Thus, full spatial diversity order is unreachable due to the rank deficiency of the keyhole effect. This result is also consistent with \cite{sanayei2007antenna}. Besides, as substantiated in \cite{shi2017asymptotic}, the time diversity order that can be achieved by Type-I HARQ is determined by the maximum number of transmissions, i.e., $K$.

\subsection{Modulation and Coding Gain}\label{Coding_gain}
The modulation and coding gain $\mathcal{C}(R)$ quantifies how much transmit power can be reduced by using a certain modulation and coding scheme (MCS) to achieve the same outage performance. In other words, $\mathcal C(R)$ characterizes how much gain can be benefited from the adopted MCS. With the asymptotic results, the explicit expression of $\mathcal{C}(R)$ can be obtained for different relationships between $N_T$ and $N_R$. %Besides, the increase of $\mathcal{C}(R)$  is conducive to the improvement of the outage performance.
To simplify our discussion, we only consider the case of $N_T=N_R$. By plugging (\ref{eqn:asy I}) into (\ref{eqn:pro}), we can obtain the modulation and coding gain in the case of $N_T=N_R$ as
\begin{equation}\label{eqn:coding_gain}
\mathcal{C}(R) = \frac{1}{{N_{T}}\left(2^R-1\right)}(N_T{({\Gamma \left( {{N_T}} \right)})}^2)^{\frac{1}{N_T}}.
\end{equation}
It is easily evident from (\ref{eqn:coding_gain}) that the modulation and coding gain is a decreasing function w.r.t. the transmission rate $R$.

\section{Numerical Analysis}\label{sec:num}
In this section, numerical results and Monte-Carlo simulations are presented for verifications and discussions. Unless otherwise specified, the system parameters are set as $R=3$ bps/Hz, $N_T=N_R=2$ and $K=3$. Moreover, we assume that the transmit SNR are identical across all HARQ rounds, i.e., $\gamma_1=\cdots=\gamma_K=\gamma$ in the sequel, and the labels ``Sim.'', ``Exa.'' and ``Asy.'' represent the simulated, the exact and the asymptotic outage probabilities, respectively.

\begin{figure}[!htb]
        \centering
        \includegraphics[width=3.5in]{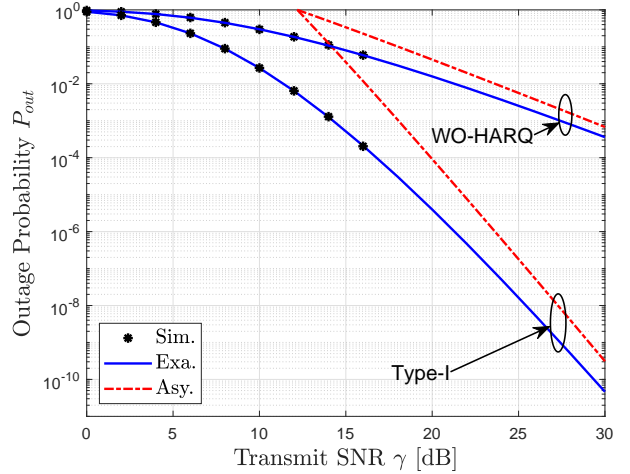}
        \caption{The outage probability $P_{out}$ versus the transmit SNR $\gamma$ with  $N_{T}=N_{R}=2$.}\label{fig:22}
\end{figure}

\begin{figure}[!htb]
    \centering
    \includegraphics[width=3.5in]{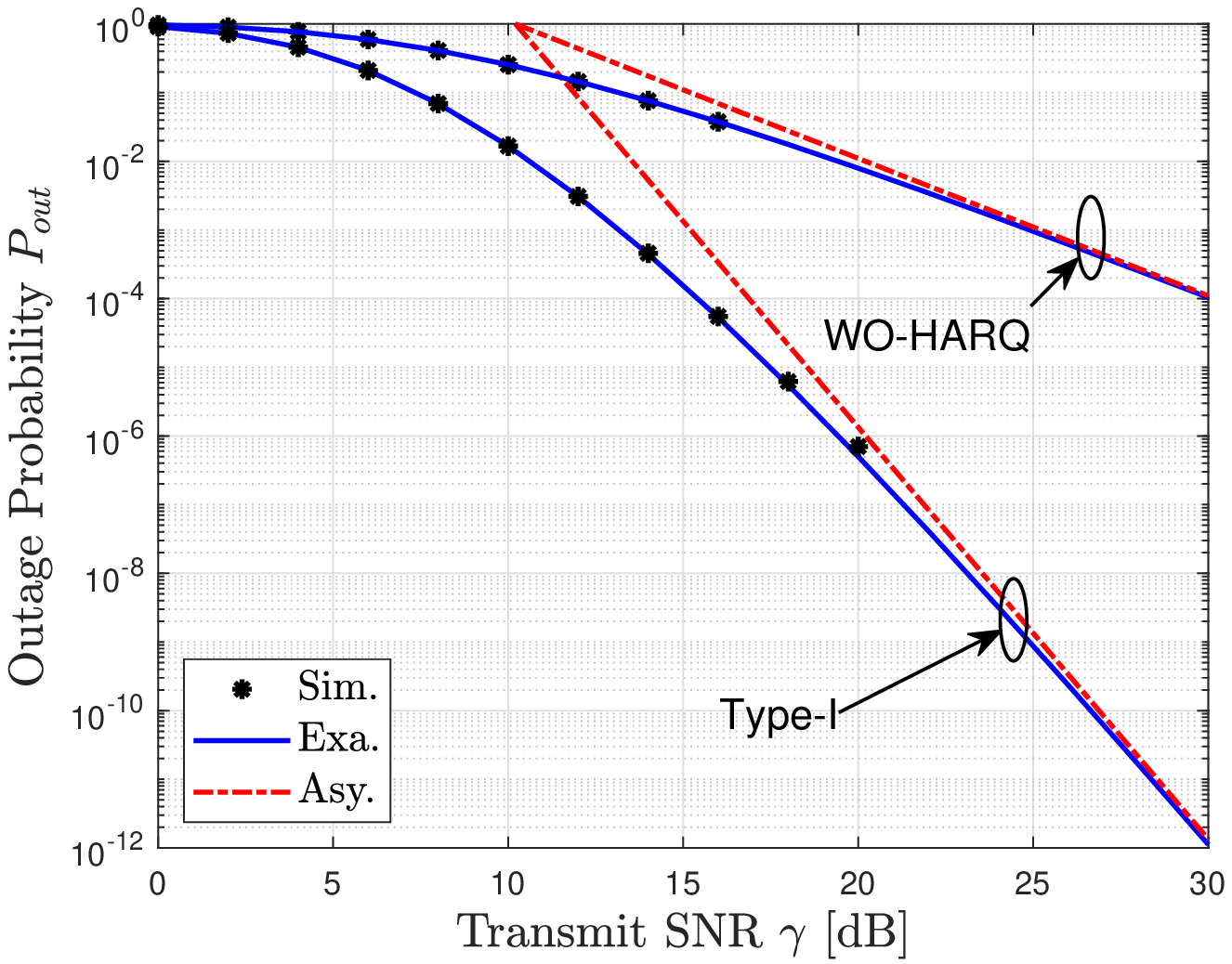}
    \caption{The outage probability $P_{out}$ versus the transmit SNR $\gamma$ with $N_{T}=3$ and $N_{R}=2$.}\label{fig:23}
\end{figure}

\begin{figure}[!htb]
    \centering
    \includegraphics[width=3.5in]{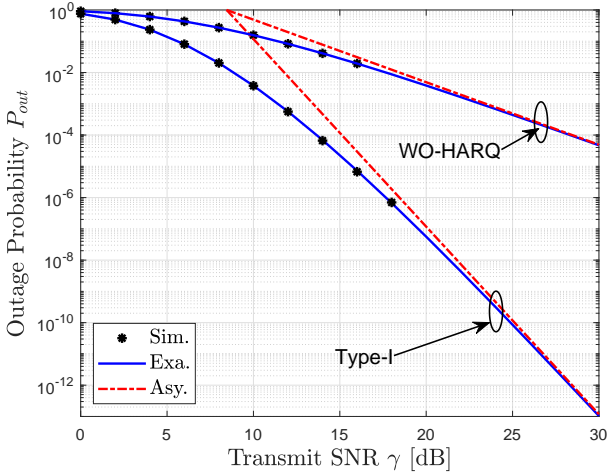}
    \caption{The outage probability $P_{out}$ versus the transmit SNR $\gamma$ with $N_{T}=2$ and $N_{R}=3$.}\label{fig:32}
\end{figure}

According to Section \ref{sec:out}, Figs. \ref{fig:22} - \ref{fig:32} show the simulated, exact and asymptotic outage probabilities of the Type-I HARQ scheme versus the transmit SNR under different relationships between the numbers of transmit and receive antennas. It is observed from Figs. \ref{fig:22} - \ref{fig:32} that the exact result match well with the simulated one for the Type-I HARQ scheme.  Moreover, the asymptotic outage curves of MIMO-HARQ schemes match very well with the exact and simulation ones at high SNR, which validates the asymptotic results as well. This observation is consistent with the asymptotic outage analysis in Section \ref{sec:out}. In addition, the slope of asymptotic outage probability is equal to $K\mathrm{min}\left(N_T,N_R\right)$, which reflects the decreasing slope of the outage curve and coincides well with the results in  Section \ref{sec:div_order}. Furthermore, by comparing to the MIMO-V2V communications without using HARQ (labeled as ``No-HARQ'' in Figs. \ref{fig:22} - \ref{fig:32}), the proposed HARQ-assisted schemes exhibit a superior performance.

\begin{figure}
    \centering
    \includegraphics[width=3.5in]{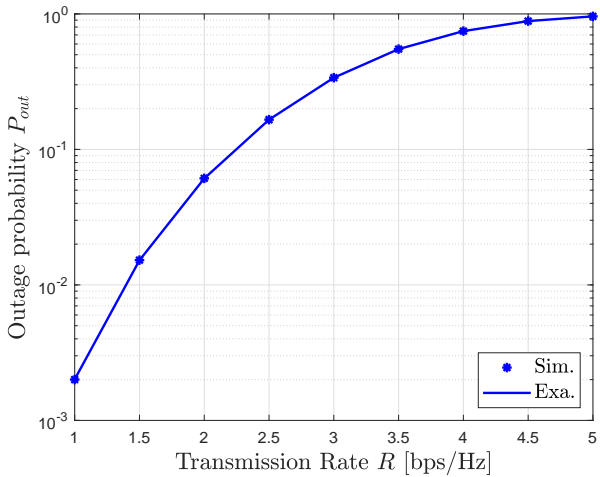}
    \caption{The outage probability $P_{out}$ versus the transmission rate $R$ by setting parameters as $\gamma=5$ dB.}\label{fig:R22}
\end{figure}

Figs. \ref{fig:R22} depicts the outage probability of the proposed systems versus the transmission rate $R$. As observed from Fig. \ref{fig:R22}, the exact outage curve coincide with the simulated outage one. Besides, it can be seen that the outage probability is an increasing function of the transmission rate $R$, this is essentially due to the tradeoff between the throughput and reliability. Hence, the transmission rate should be properly chosen in practical V2V communication systems. Fortunately, the increasing monotonicity and convexity of the asymptotic outage probability with respect to $R$ can greatly ease the optimal rate selection.

\begin{figure}
    \centering
    \includegraphics[width=3.5in]{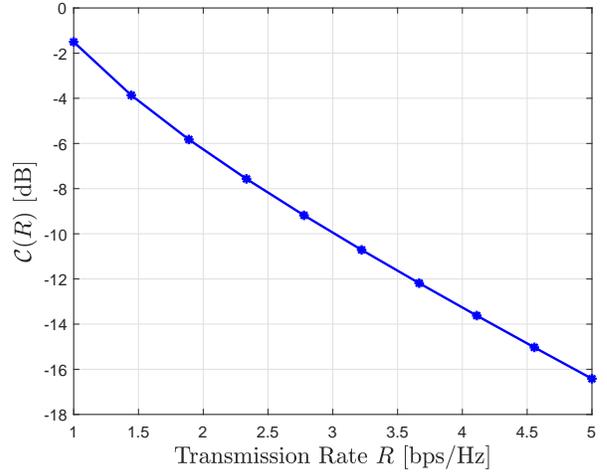}
    \caption{The modulation and coding gain $\mathcal{C}(R)$ versus the transmission rate $R$ by setting parameters as $N_{T}=N_{R}=2$ and $K=3$.}\label{fig:R23}
\end{figure}

Additionally, Fig. \ref{fig:R23} shows the impacts of the transmission rate $R$  on the modulation and coding gain $\mathcal{C}(R)$ for the proposed systems. As expected, the modulation and coding gain decreases with the increase of the transmission rate. The observations in Fig. \ref{fig:R23}  are consistent with the asymptotic analysis in Section \ref{Coding_gain}.

\section{Conclusions}\label{sec:con}
In this paper, we have investigated the outage performance for MIMO-HARQ assisted V2V communications with keyhole effect. To compensate the performance degradation caused by the rank-deficiency of keyhole effect, the Type-I HARQ has been employed to boost the transmission reliability. With the help of  the Meijer G-function, we have derived the exact outage expression. Moreover, the asymptotic outage analyses has been conducted to obtain meaningful insights. In particular, we have derived the diversity order and modulation and coding gain of the proposed systems. However, it has been shown that only full time diversity order can be achieved, while full spatial diversity order is unreachable as compared to MIMO-HARQ systems without keyhole effect. Additionally, the monotonically increasing and convex function properties of asymptotic outage probability w.r.t. transmission rate was obtained in this paper, which provides a convenient way to select a suitable rate for efficiency and reliability tradeoff. In the end, the Monte Carlo simulations have validated the analytical outcomes.

\appendices

\bibliographystyle{ieeetran}
\bibliography{manuscript_1}

\end{document}